# The Third Visual Pathway for Social Perception

David Pitcher

Department of Psychology, University of York, Heslington, York, YO105DD, U.K.

Corresponding author:   David Pitcher: david.pitcher@york.ac.uk

Department of Psychology, University of York, Heslington, York, YO10 5DD, U.K

## Abstract

Influential models of primate visual cortex describe two functionally distinct pathways: a ventral pathway for object recognition and the dorsal pathway for spatial and action processing. However, recent human and non-human primate research suggests the existence of a third visual pathway projecting from early visual cortex through the motion-selective area V5/MT into the superior temporal sulcus (STS). Here we integrate anatomical, neuroimaging, and neuropsychological evidence demonstrating that this pathway specializes in processing dynamic social cues such as facial expressions, eye gaze, and body movements. This third pathway supports social perception by computing the actions and intentions of other people. These findings enhance our understanding of visual cortical organization and highlight the STS's critical role in social cognition, suggesting that visual processing encompasses a dedicated neural circuit for interpreting socially relevant motion and behavior.

**Visual pathways inform models of cognition and behavior**

Explaining the neural processes that enable humans to see, understand, and interact with the people, places, and objects in our environment is a fundamental aim of psychology. A particularly fruitful theoretical approach in pursuit of this goal has been to deduce the cognitive operations performed by specific brain areas based, at least in part, on their anatomical connectivity. Models of cognition informed by neuroanatomical data reveal how complex cognitive functions are constructed through the integration of information from primary sensory brain areas. For example, influential models of the visual cortex propose that it contains two functionally distinct pathways: a ventral pathway for visual object recognition and a dorsal pathway for performing visually guided physical actions (Ungerleider and Mishkin, 1982, Milner and Goodale, 1986). The two visual pathway model transformed our understanding of cortical organisation and acted as the blueprint for the subsequent mapping of visual functions in the primate brain (Kravitz et al., 2011; Kravitz et al., 2013; Milner & Goodale, 1986; Ungerleider & Mishkin, 1982).

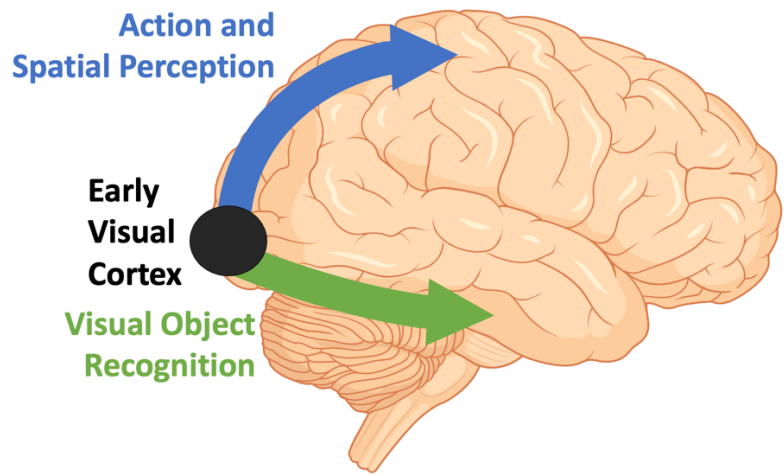

**Figure 1. The two visual pathway model (Milner & Goodale, 1986; Ungerleider & Mishkin, 1982). The ventral pathway (green) is used for visual object recognition. The dorsal pathway (blue) is used for performing visually guided physical actions and locating objects in visual space.**

Characterizing the visual system as functional pathways enables researchers to describe the cognitive operations of the brain at a level that can encompass different species (Bruce et al., 1981; Freiwald et al., 2016; Gross et al., 1972; Miller et al., 2016; Perrett et al., 1992; Tsao et al., 2008), different stages of development (Grossmann, 2021; Im et al., 2025) and different experimental methods. These include: tracer and tractography studies that map anatomical connectivity (Gschwind et al., 2012; Seltzer & Pandya, 1978; Ungerleider & Desimone, 1986a); physiology and neuroimaging studies that map the neural response at different spatial scales (Bruce et al., 1981; Hadj-Bouziane et al., 2012; Isik et al., 2017; Saxe et al., 2004; Yan et al., 2025); neuropsychological and brain stimulation studies that causally define the behavioural functions performed in disrupted brain areas (Campbell et al., 1990; Fried et al., 1982; Pitcher, 2014; Prabhakar et al., 2025); developmental studies that reveal how visual cognition develops in infants (Deen et al., 2017; Livingstone et al., 2017; Lloyd-Fox et al., 2009). Models of cognition informed by brain anatomy create a common framework that enables scientists to integrate empirical data using any of these methods. This facilitates understanding between those who study the brain at the behavioural, cognitive and neural levels.

**A Social Gap: The need for the third pathway**

Despite the ubiquity and success of the two visual pathway model there were some obvious omissions in explaining the broad scope of human behaviour. Perhaps most notably, the focus on object recognition and physical action ignored the visual mechanisms that underpin visual social cognition. Humans are social primates living in a world of limited resources. This makes understanding the behaviour and intentions of the people around us a priority. To resolve this gap, we proposed a major revision of the original two pathway model in 2021 to include a third visual pathway specialised for social perception (Pitcher, 2021; Pitcher & Ungerleider, 2021). Social perception is predicated on visually analysing and understanding the actions of others and responding appropriately. One region of the brain in particular, the superior temporal sulcus (STS), computes the sensory information that facilitates these processes (Allison et al., 2000; Kilner, 2011; Perrett et al., 1992). The STS selectively responds to moving biological stimuli (e.g., faces and bodies) and computes the visual social cues that help us understand and interpret the actions of other people. These include facial expressions, eye gaze, body movements and the audio-visual integration of speech (Calder et al., 2007; Calvert & Campbell, 2003; Grossman & Blake, 2002; Pitcher et al., 2020; Puce et al., 1998). The third visual pathway projects from primary visual cortex into the STS, via the motion-selective area V5/MT (Figure 2) (Pitcher & Ungerleider, 2021). In this chapter, I will argue that visual motion, particularly biological motion, forms the basis of the third pathway's role in enabling the perception and understanding of others' behavior.

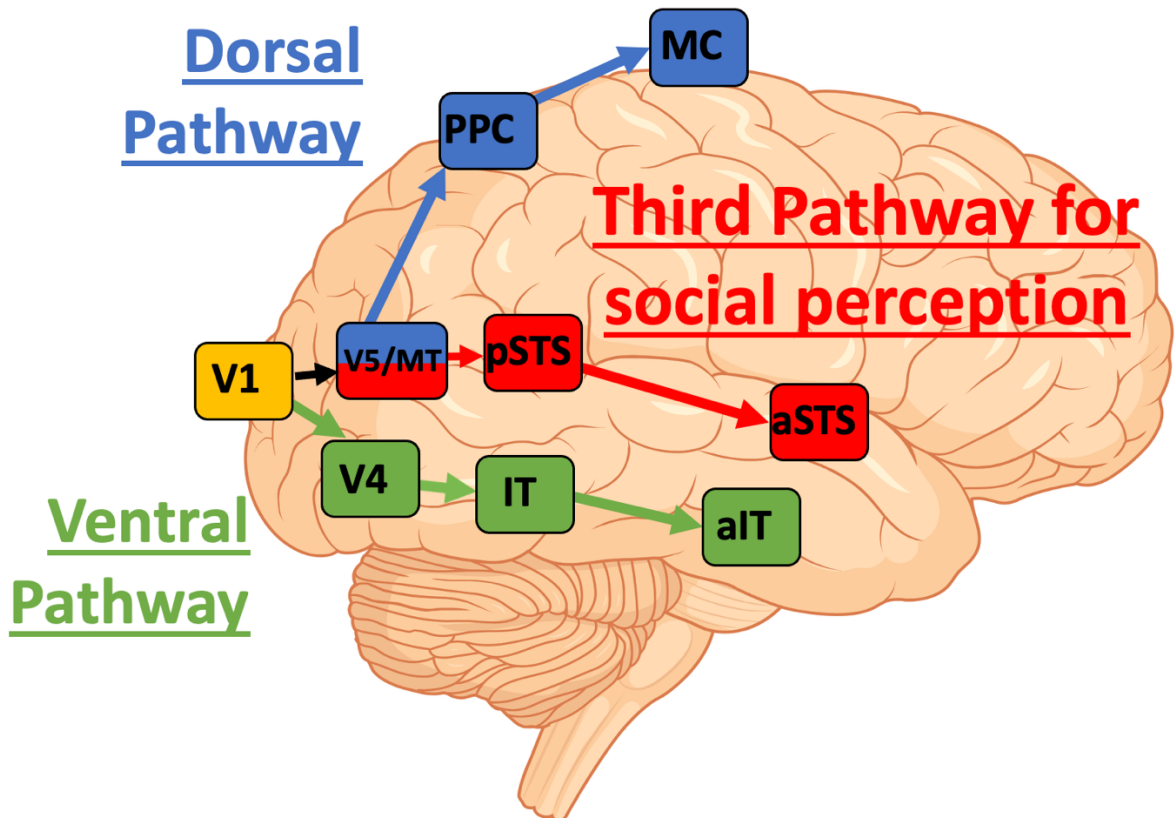

**Figure 2. The revised visual pathway model (Pitcher & Ungerleider, 2021). The ventral pathway (green) is used for visual object recognition. The dorsal pathway (blue) is used for performing visually guided physical actions and locating objects in visual space. The third visual pathway (red) for social perception is used to understand the social behaviour of other people. PPC – Posterior Parietal Cortex; MC – Motor Cortex; pSTS – Posterior Superior Temporal Sulcus; aSTS – Anterior Superior Temporal Sulcus; IT - Inferior Temporal Cortex; aIT – Anterior Inferior Temporal Cortex**

The third visual pathway re-conceptualised one of the dominant models of primate visual cortex by addressing a key omission, inherent in the original theory. Namely, what are the functional and structural connections between early visual cortex and the STS in the human brain. The STS is engaged in both social perception and in the more complex cognitive computations that facilitate social interaction (Allison et al., 2000; Hein & Knight, 2008; Kilner, 2011; Perrett et al., 1992). However, the connectivity between early visual cortex and the STS remains poorly characterised. This led some researchers to view the STS as an extension of the ventral pathway, rather than as a functionally and anatomically independent pathway in its own right. For example, models of face processing propose that all facial aspects (e.g., identity and expression recognition) are processed using the same early visual mechanisms (Bruce &

Young, 1986; Haxby et al., 2000; Pitcher, Walsh, et al., 2011) before diverging at higher levels of processing, rather than as dissociable processes that begin in early visual cortex.

The role of the STS in social cognition is well established (Allison et al., 2000; Perrett et al., 1992). The best evidence comes from the extensive literature that has demonstrated how the STS responds to a wide variety of social cues. The information used by primates to calculate the meanings and intentions of others is generated by their actions (Puce, 2024). These actions can be generated by a face, body, speech and sound. fMRI studies of the STS have shown that it contains regions that selectively respond to different types of visual and auditory stimuli. These include faces (Calder et al., 2007; Phillips et al., 1997; Pinsk et al., 2009), bodies (Beauchamp et al., 2003; Saxe et al., 2004), point-light walkers (Grossman & Blake, 2002), the human voice (Belin, 2006), language (Binder et al., 1997) and the audio-visual integration of speech (Calvert & Campbell, 2003). In addition, the temporoparietal junction (TPJ) (an adjacent brain area posterior and superior to the STS) responds to theory of mind tasks (Saxe & Kanwisher, 2003) in which participants are required to interpret the actions of characters in brief stories. An fMRI study that simultaneously mapped the responses to all these types of stimuli along the STS identified regions that selectively responded to certain types of social input as well as regions that responded to multiple contrasts (Deen et al., 2015). This study demonstrated that while the posterior regions are heavily involved in theory of mind and biological motion, more anterior regions are specialized for speech processing. The pSTS acts as a bridge to the "mentalizing" network, specifically the Temporoparietal Junction (TPJ). By identifying the direction of gaze or the target of a physical action, the STS provides the perceptual data necessary for the brain to infer another person's intentions, beliefs, or desires. This proximity of brain areas computing multi-sensory information relevant to social interaction functionally dissociates the STS from the established roles of the ventral and dorsal visual pathways.

## **Neuroanatomy of the third pathway**

The cognitive operations performed in a particular brain area can be deduced (at least partially) by the anatomical connectivity of that area. The most compelling evidence for a cortical pathway into the STS that bypasses the ventral pathway comes from non-human primate neuroanatomy. Many of these studies have been conducted in macaque monkeys, a highly social non-human primate. Neuroanatomical studies have identified a direct cortical projection extending from the primary visual cortex (V1) to the motion-selective area MT (Figure 3) (Shipp & Zeki, 1989; Ungerleider & Desimone, 1986a). Importantly, this pathway is anatomically segregated from cortical connections between V1, V2 and V4 that project directly into the inferior temporal cortex of the ventral pathway (Ungerleider & Desimone, 1986b). Anterior connectivity from V5 / MT feeds into adjacent anterior motion-selective areas, specifically MST and FST (Boussaoud et al., 1990) with FST ultimately projecting into the dorsal bank and fundus of the STS. Crucially, this lateral trajectory is anatomically segregated from the connections along the ventral brain surface (Boussaoud et al., 1991).

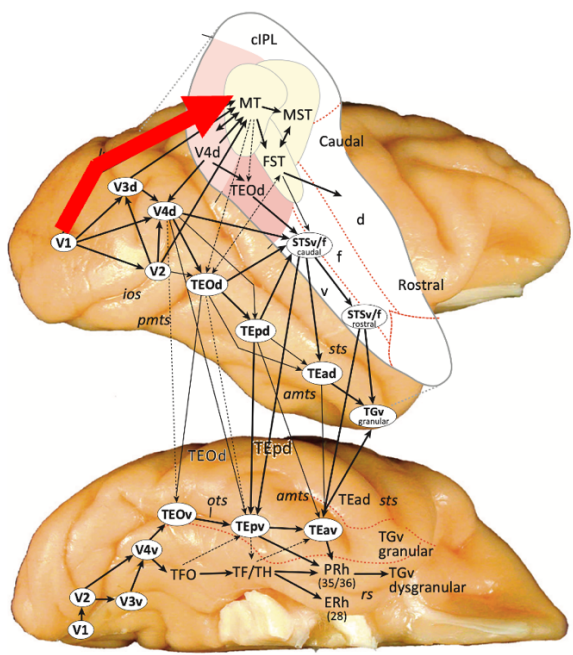

**Figure 3: The direct cortical connection from V1 to V5/MT (red) in macaque cortex that dissociates the third and dorsal pathways from the ventral pathway**

While the macaque data acts as a roadmap for mapping homologies across species, such invasive anatomical studies are not possible in humans. Non-invasive neuroimaging techniques such as diffusion tensor imaging (DTI) can be used to map the anatomical connectivity of the STS. Tractography studies have identified a white matter pathway along the STS that is anatomically segregated from white matter pathways on the ventral surface (Gschwind et al., 2012; Makris et al., 2013). Other studies have combined tractography with other methods to further characterize the anatomical connectivity of the occipital cortex to the STS. Babo-Rebelo and colleagues combined intracranial electroencephalography (iEEG) from epilepsy patients with white matter tractography from healthy adults to investigate how the human brain processes faces (Babo-Rebelo et al., 2022). Their results demonstrated that while the inferior occipital cortex (part of the ventral visual pathway) was the primary entry point of the face processing network there was also evidence of direct connections into the STS for processing the dynamic aspects of faces such as eye gaze direction.

A more recent study that combined data from neuropsychological patients with DTI data from healthy control participants also provided evidence consistent with the third visual pathway. Prabhakar and colleagues tested 108 patients with focal brain lesions: some lesions encompassed the STS (but not the ventral pathway), while others encompassed the ventral pathway (but not the STS) (Prabhakar et al., 2025). Their results showed that patients with STS damage were more impaired at recognizing moving facial expressions than static ones, whereas patients with lesions to the ventral pathway were more impaired at recognizing static facial expressions than moving ones. Lesion locations were established from structural brain scans and then mapped to the white matter tracts that would typically connect these areas in the undamaged brain. The results showed that STS lesions producing impairments in recognizing moving expressions were anatomically dissociable from ventral lesions producing impairments in recognizing static expressions. STS lesions were associated with the right middle longitudinal fasciculus and arcuate fasciculus, while ventral lesions were associated with the inferior longitudinal fasciculus and inferior fronto-occipital fasciculus. This result elegantly demonstrated that the behavioral impairments in facial expression recognition associated with these distinct lesion patterns are consistent with the third visual pathway (Pitcher, 2021, 2022, 2023, 2025; Pitcher & Ungerleider, 2021).

DTI studies of typical participants further demonstrate how the STS is connected to temporal, parietal, frontal, and occipital regions by several major long-range white matter tracts. The middle longitudinal fasciculus (MdLF) projects longitudinally through the superior temporal lobe connecting the STS with the inferior parietal lobule and angular gyrus and extrastriate areas of the occipital cortex, playing a pivotal role in integrating visual and auditory information with visuospatial attention and language networks (Makris et al., 2013; Wang et al., 2013). The posterior STS (pSTS) is connected to the frontal gyrus (via the dorsal visual pathway) by the arcuate fasciculus (AF) and superior longitudinal fasciculus (SLF) which supports phonological processing (Catani et al., 2005) and biological motion perception (Rilling et al., 2008). The anterior and mid-portions of the STS are integral to the ventral stream, connecting to the ventral inferior frontal gyrus via the Extreme Capsule (EmC) and Inferior Fronto-Occipital Fasciculus (IFOF). This pathway is critical for semantic processing, allowing the brain to attach meaning to auditory and visual inputs (Saur et al., 2008). The anterior STS (approaching the temporal pole) is connected to the orbitofrontal cortex and the limbic system via the Uncinate Fasciculus (UF). This connection facilitates the integration of social stimuli with emotional valuation and memory, a process vital for social reasoning (Von Der Heide et al., 2013).

## A full representation of the visual field: Dissociating the ventral pathway from the third pathway

The primate visual system can be mapped and at least partially understood by mapping the retinotopic organization. This is because specific regions of the visual cortex process information from specific spatial locations in the external world as projected onto the retina. A fundamental principle of this organization in early and intermediate visual areas is contralateral visual field bias. The left hemisphere preferentially processes the right visual field, while the right hemisphere preferentially processes the left visual field. Visual field mapping studies are an experimental method used to infer the structural connectivity of the brain by analyzing how visual space is represented across the cortex. The principle is that anatomically interconnected brain areas typically share similar topographic representations of the visual field. However, converging evidence from recent neuroimaging and stimulation studies suggests that this rule does not apply uniformly across the high-level visual cortex. Specifically, we demonstrated that there is a functional dissociation between the face-selective regions of the ventral pathway (implicated in identity recognition) and the third pathway (implicated in dynamic social cognition) (Pitcher et al., 2020).

Prior to the advent of non-invasive neuroimaging techniques like fMRI, the mapping of visual field responsiveness was established through direct electrophysiological recording of neuronal activity in non-human primates. Visual field mapping studies confirm that these areas also represent the same specific sectors of the visual field, specifically exhibiting a strict bias for the contralateral (opposite) side. For example, in the macaque visual system, areas V1, V2, and V4 are known to have dense anatomical interconnections (Gattass et al., 1997). This combination of visual maps with neuroanatomical studies can be used to strengthen the evidence for cortico-cortical connectivity because if regions share a visual topography, then they are likely to be part of the same visual pathway. One study reported a functional gradient along the motion-selective cortex of the macaque that projects into the STS (Desimone & Ungerleider, 1986). The middle temporal area (MT) is driven primarily by input from the contralateral visual field. However, moving anteriorly along the sulcus reveals a progressive expansion of the receptive fields into areas MST (Medial Superior Temporal) and FST (Fundus of the Superior Temporal) begin to represent significantly larger portions of the ipsilateral visual field. Subsequent neuroimaging studies in humans have mirrored these phylogenetic findings; anterior sub-regions of the motion complex (hMT+) exhibit a greater proportion of ipsilateral representation compared to their posterior counterparts (Amano et al., 2009; Huk et al., 2002). Collectively, this converging evidence from single-unit physiology and human neuroimaging delineates a specialized processing stream in the lateral brain, one that rapidly transcends the contralateral biases of early visual inputs to construct a global, unified representation of visual motion that progresses into the STS.

A seminal early visual mapping study in macaques demonstrated that neurons in the STS respond to stimuli shown in almost the entirety the visual field (Bruce et al., 1981). The authors recorded activity from neurons in the dorsal bank and fundus of the anterior portion of the STS to a range of different stimuli. Almost all neurons responded to visual stimuli (including faces) and consistent with later fMRI studies in humans over responded to auditory and somatosensory stimuli (or both). The receptive fields were large and mostly bilateral typically covering both visual hemifields and often the entire visual field. This dissociated the sampled region in the STS from areas of the ventral processing pathway for object recognition. Furthermore, the neurons showed a strong preference for moving stimuli over stationary ones. Many cells were directionally selective, responding best to motion in specific directions. This led the authors to conclude that the STS plays a specialized role in monitoring the global sensory environment, integrating visual motion with other modalities to track external events. It is notable that this study characterized the full visual field response, the preference for motion, the polysensory response and the face-selectivity in the STS in a single study. This was at least a decade prior to these conclusions being made in other species and using other experimental methods.

The STS shows a greater response for moving more than static stimuli in the STS (Fox et al., 2009; Pitcher, Dilks, et al., 2011) and the motion-selective area V5/MT is core component of the third visual pathway. This suggests that moving visual stimuli should be used to map the visual field biases. To establish the visual field response in the human STS we performed an fMRI study in which we presented short videos of actors posing different facial expressions (Pitcher et al., 2020). Prior evidence had established that the occipital face area (OFA) (Gauthier et al., 2000) and the fusiform face area (FFA) (Kanwisher et al., 1997; McCarthy et al., 1997) both exhibit a contralateral visual field bias (Hemond et al., 2007; Kay et al., 2015). These studies demonstrated that a visual stimulus, such as a static face, appears in the left visual field, it elicits a robust response in the right FFA and OFA, but a negligible response in their left-hemisphere counterparts. This organization facilitates the detailed analysis of invariant facial features required for identification, a process that arguably benefits from the retinotopic separation of visual information. This is consistent with established models of the ventral visual pathway which exhibits a posterior to anterior gradient that increasingly responds to the ipsilateral visual field (Grill-Spector & Malach, 2004).

To establish the visual field response in the STS we performed a neuroimaging study designed to maximize functional responses in the STS using dynamic stimuli (Pitcher et al., 2020). Participants viewed videos of actors posing various facial expressions presented in different quadrants of the visual field. The results revealed a striking dissociation. While the motion-selective area V5/MT and the ventral face areas (FFA, OFA) exhibited the expected contralateral bias, the face-selective region in the posterior STS (pSTS) did not (Figure 4). The pSTS responded with equal magnitude to dynamic social signals regardless of whether they appeared in the left or right hemifield, or the upper and lower parts of the visual field. We further demonstrated that the bilateral visual response in the STS was behaviourally relevant in a TMS study. TMS delivered over the right OFA impaired expression recognition in the contralateral field only, while TMS delivered over the right STS impaired expression recognition in both visual fields (Pitcher et al., 2020). Finally, we performed an fMRI study of a prosopagnosic patients with a lesion encompassing the right FFA and right OFA only (Sliwinska, Bearpark, et al., 2020). Despite this damage there was a still a bilateral visual field response in the STS thereby demonstrating independent functional inputs into the STS that bypass the ventral visual pathway. This finding was further corroborated by Finzi and colleagues using preferred retinal focus (pRF) mapping with cartoon faces, which confirmed that the receptive fields in the STS are significantly larger and more bilateral than those in the ventral stream (Finzi et al., 2021). This study also investigated structural connections between early visual cortex and face-selective regions using diffusion MRI. They found that ventral face-selective regions had a higher proportion of white matter tracts connecting to central eccentricities of early visual cortex, while lateral regions had a more uniform distribution of connections across eccentricities. This structural difference mirrored the functional differences observed in pRF properties and VFC.

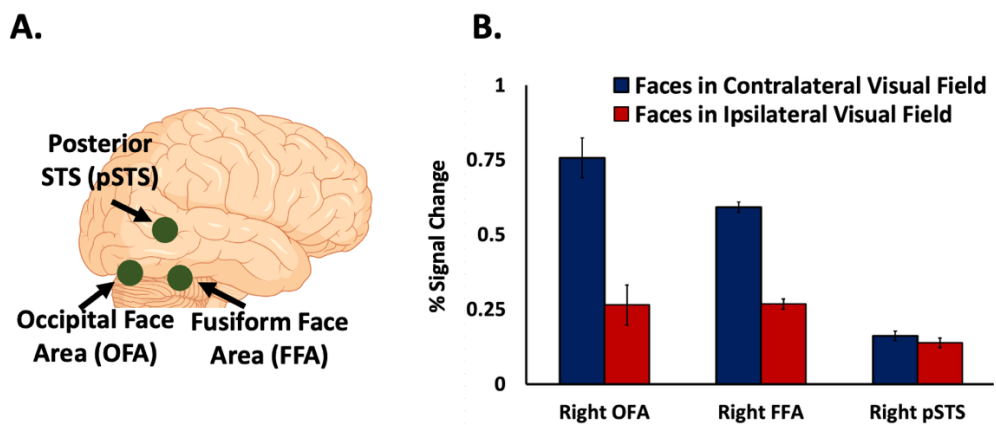

**Figure 4a. The face-selective areas in occipitotemporal cortex. These include the posterior STS (pSTS), the fusiform face area (FFA) and occipital face area (OFA). Figure 4b. Visual field responses in face-selective areas. The right OFA and FFA show a contralateral bias while the pSTS has no visual field bias.**

The absence of a contralateral bias in the STS implies a functionally distinct underlying neural architecture. Because the retinae project contralaterally to the early visual cortex, the ipsilateral responses observed in the pSTS can only arise via transfer from the opposing hemisphere. This suggests that the lateral visual pathway possesses a significantly higher degree of functional interhemispheric connectivity, via the corpus callosum, than in the ventral pathway. We argued that the panoramic visual processing in the STS is entirely consistent with the ecological and social demands placed on primates (Pitcher & Ungerleider, 2021). Unlike object recognition or reading, which often require foveating on a specific, stationary target to resolve fine detail, social interaction is inherently dynamic and spatially distributed. Social cognition rarely occurs in a dyadic vacuum; it is sometimes conducted between an individual and a group. To successfully navigate a social environment, we need to compute the locations, intentions, and movements of multiple biological organisms simultaneously across the entire visual field. For example, while attending to a central speaker amongst a larger group, we need to remain sensitive to social cues in the periphery and shift our attention as required. If the pSTS were restricted by a contralateral bias, the integration of these disparate social signals may take longer. By processing the entire visual field, we benefit from a continuous, holistic representation of any biological organisms in the social environment.

**Motion equals meaning: Dynamic face processing in the third pathway**

Faces are rich sources of social information that simultaneously convey someone's identity, attentional focus, and emotional state. Our daily lives are dependent on swiftly and accurately decoding this socially relevant information which is the basis of understanding, empathy and social interaction. Cognitive and neurobiological models of face processing (Allison et al., 2000; Bruce & Young, 1986; Haxby et al., 2000) propose that these cues can be broadly segregated into cues that rely on invariant aspects (e.g. facial structure and shape) and changeable aspects (e.g. eye, mouth and head movements). The invariant (or static) cues contribute to identity discrimination (Grill-Spector et al., 2004; Megreya & Burton, 2006; Pitcher, Caulfield, et al., 2023; Sliwinska et al., 2022) while the changeable aspects contribute to facial expression discrimination (Krumhuber et al., 2023; Pitcher et al., 2020). This functional dissociation between static and dynamic facial cues is supported by evidence from different methods. These include behavioural experiments (Calvo et al., 2016; Dobs et al., 2018; Lander et al., 1999), neuropsychology (Prabhakar et al., 2025; Sliwinska, Bearpark, et al., 2020), neuroimaging (Fox et al., 2009; Kucuk et al., 2024; Puce et al., 1998; Schultz & Pilz, 2009; Yan et al., 2025) and developmental studies (Im et al., 2025; Kosakowski et al., 2022). Our proposal of the third visual pathway proposes that is changeable aspects of a face that are processed in the STS and that these cues are used to support the dynamic aspects of social cognition.

Neuropsychological studies of patients with discrete cortical lesions offer a unique way to causally establish the structural and functional connectivity of the lateral pathway in the human brain. As early as 1984, Bauer (Bauer, 1984) postulated the existence of a direct pathway into the STS based on his study of prosopagnosic patient GKT. His patient exhibited different galvanic skin responses to familiar and unfamiliar faces even when he was unable to consciously recognize the faces. This suggested alternate cortical pathways for different aspects of face processing, but the lack of structural brain imaging made the theory speculative (Figure 5a). The advent of functional brain imaging has since been used to report studies of multiple prosopagnosic patients who exhibit a face-selective response in the STS, despite having lesions encompassing the brain area in which the FFA and OFA are typically located (Dalrymple et al., 2011; Gao et al., 2019; Rezlescu et al., 2014; Rezlescu et al., 2012; Steeves et al., 2006). These patterns of activation demonstrate that face-selective activity in the STS can occur even when face-selective areas in the ventral pathway have been damaged or destroyed.

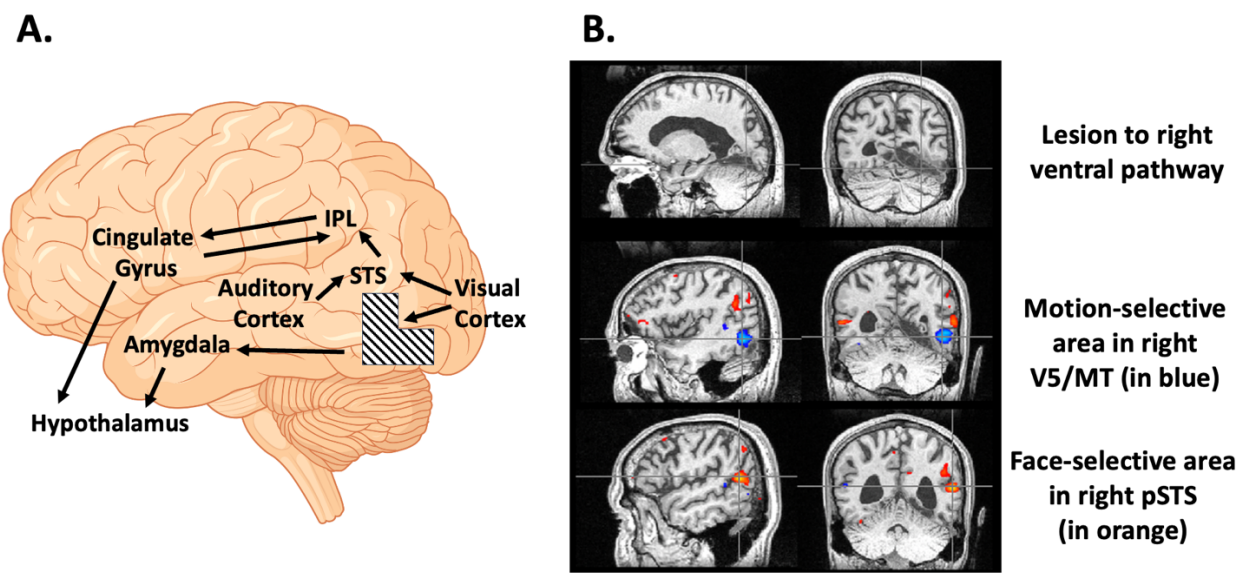

**Figure 5a. An adaptation of the schematic representation taken from Bauer (1984). The blocked area predicts the location of the lesion in Patient GSK (Bauer, 1984). Note that the author predicted a pathway from visual cortex into the STS prior to the advent of human functional neuroimaging. Figure 5b. Data from prosopagnosic patient Herschel (Rezlescu et al., 2014; Rezlescu et al., 2012; Sliwinska, Elson, et al., 2020). Despite an extensive right occipitotemporal lesion Herschel still has functional activation to motion in V5/MT and moving faces in the pSTS.**

These results are seemingly inconsistent with established face processing models that stipulated a single-entry point for the face processing network (Bruce & Young, 1986; Haxby et al., 2000; Pitcher, Walsh, et al., 2011). This early structural encoding stage is located in the in the inferior occipital gyrus, also called the occipital face area (OFA) (Gauthier et al., 2000; Pitcher et al., 2009; Pitcher et al., 2007). An alternative model was offered by O'Toole & colleagues (O'Toole et al., 2002) who instead proposed that moving faces are processed via a pathway that runs from early visual cortex into the STS via V5/MT. We tested the functional connections proposed in this model in a recent fMRI study of Herschel, a prosopagnosic patient with a right ventral occipitotemporal lesion (Sliwinska, Bearpark, et al., 2020). Across three experiments we measured the neural response to moving and static faces, and to face videos shown in the two visual fields across face-selective areas. Results showed the response to moving and static faces in the patients right pSTS was not significantly different from control participants (Figure 5a). We also observed no visual field bias in his right pSTS despite his right ventral occipitotemporal lesion. This pattern of results is consistent with the existence of a direct cortical pathway from early visual cortex into the STS and further suggests this pathway may have a greater degree of interhemispheric connectivity than the ventral stream.

The crucial role of motion on the face-selective response in the STS has also been demonstrated in neurologically normal experimental participants. Neuroimaging studies demonstrate that the face-selective area in the posterior STS (pSTS) exhibits a greater response to moving, more than static faces, while ventral face-selective regions, like the fusiform face area (FFA) and occipital face area (OFA) show little (or no) preference for dynamic over static faces (Fox et al., 2009; Kucuk et al., 2024; LaBar et al., 2003; Nikel et al., 2022; Pitcher et al., 2019; Puce et al., 1998; Schultz & Pilz, 2009). In order to further elucidate the role of motion in the lateral pathway we combined TMS with fMRI to disrupt the pathways that support dynamic face processing in neurologically normal participants. TMS was delivered over the right OFA, or right pSTS while participants were scanned with fMRI while viewing moving or static face images (Pitcher et al., 2014). Results showed that disruption of the right OFA reduced the neural response to both static and moving faces in the right FFA. This result for static faces was subsequently replicated in a follow-up combined TMS and fMRI study of the right OFA (Groen et al., 2021). In contrast, the response to dynamic and static faces was doubly dissociated in the right pSTS. Namely, disruption of the right OFA reduced the response to static but not moving faces, while disruption of the rpSTS itself reduced the response to moving but not static faces. This demonstration that disrupting the ventral pathway had no effect on moving faces in the STS is consistent with the lateral pathway being specialised for dynamic face processing. Subsequent combined TMS / fMRI studies have used moving face videos and resting-state fMRI to further map the functional connectivity of the pSTS with both the anterior STS and amygdala (Pitcher et al., 2017) and the wider face network (Handwerker et al., 2020).

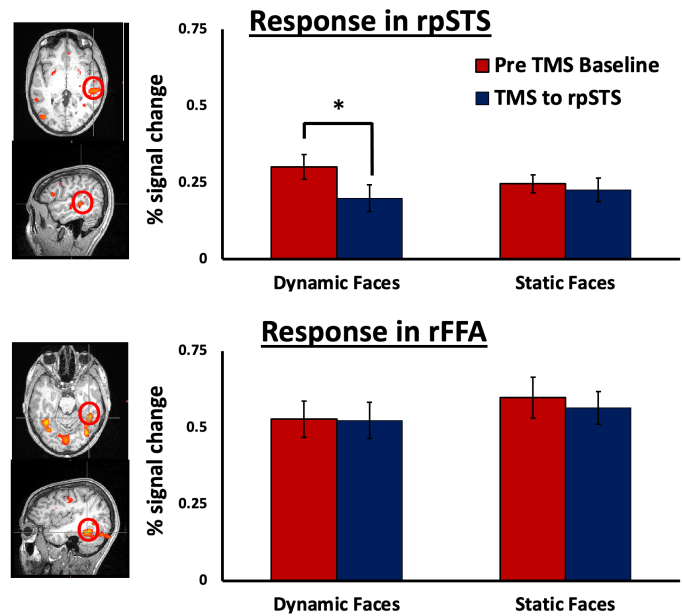

**Figure 6: TBS delivered over the face-selective pSTS reduces the BOLD response to dynamic faces, but not static faces (Pitcher et al., 2014).**

## Specialized Face Patches in the Macaque STS and Cross-Species Comparisons

Non-human primate studies, predominantly utilizing macaque monkeys, are fundamental for neuroscientific research into visual perception. These animal models are indispensable because they allow researchers to employ invasive experimental methods, such as single-unit electrophysiology and tracer injection studies, that are ethically or practically impossible in human subjects. For over fifty years, the Superior Temporal Sulcus (STS) has been a focal point of this research. Early seminal work established that neurons in the macaque STS respond selectively to visual images of faces (Bruce et al., 1981; Gross et al., 1969; Perrett et al., 1992; Perrett et al., 1982). Building on these single-cell foundations, modern functional magnetic resonance imaging (fMRI) has revealed a highly organized system, identifying at least six discrete face-selective patches spanning the posterior-to-anterior length of the STS (Tsao et al., 2003; Tsao et al., 2006). Consequently, the specific functions and interconnectivity of these areas have been mapped with increasing precision (Afraz et al., 2015; Freiwald et al., 2009; Moeller et al., 2009; Tsao et al., 2008).

A critical breakthrough in understanding how these patches operate came from Clark and Freiwald who demonstrated that the six identified patches are not functionally identical but are dissociated into two distinct processing pathways (Fisher & Freiwald, 2015). By scanning macaques while they viewed a range of both moving and static face stimuli, the researchers observed a clear topographic split. Face patches located on the dorsal bank of the STS exhibited a selective response to faces in natural motion, processing dynamic social signals. In contrast, face patches on the ventral bank selectively responded to the static structure of faces, contributing to identity recognition. Interestingly, the face patch MF, located in the fundus of the STS, exhibited a dual response to both moving and static faces, possibly serving as a functional bridge between the two banks. This dissociation suggests a face processing architecture where one pathway processes structural form (identity) and another processes changeable aspects (intent and movement). This results is consistent with the third pathway model (Pitcher & Ungerleider, 2021), suggesting a conserved evolutionary mechanism for separating cognitive mechanism for identifying an individual from how they are feeling and calculating their intentions.

A major question arises regarding hemispheric lateralization. In humans, converging evidence from fMRI, lesion studies, and TMS demonstrates that face processing is preferentially lateralized to the right hemisphere, particularly within the posterior STS (pSTS) (Kucuk et al., 2024; Pitcher, Dilks, et al., 2011; Sliwinska, Elson, et al., 2020; Sliwinska & Pitcher, 2018). De Winter et al. (2015) directly addressed this interspecies difference (De Winter et al., 2015). Human and macaque subjects were scanned viewing moving face stimuli, humans exhibited a strong right-lateralized bias, whereas macaques showed no such asymmetry. The authors speculated that this difference is likely driven by the evolution of the human brain's left hemisphere for spoken language. As the left hemisphere became specialized for verbal processing in humans, visual face processing may have been "crowded out" to the right hemisphere, an evolutionary pressure absent in macaques. This hypothesis highlights the difficulty of assuming direct homology between human and macaque brain organization without accounting for evolutionary shifts in cortical organization.

This is further demonstrated by the anatomical discrepancy between the two species. Human face areas are found on both the ventral brain surface (e.g., the Fusiform Face Area) and the lateral surface (STS). In contrast, macaque fMRI studies predominantly locate face patches only on the lateral surface within the STS (Tsao et al., 2003; Tsao et al., 2006). Recording studies have identified face cells on the ventral brain surface (Ku et al., 2011) and future studies may find these cells can cluster into patches. Another possible explanation is technical. The mouth and jaw muscles in macaques causes signal drop out that substantially impairs the fMRI data that can be recorded from the ventral brain surface. Although it should be noted that scene-selective patches have been identified on the ventral surface (Kornblith et al., 2013).

## Moving faces beyond the posterior STS: Anterior connectivity of the third pathway

Human neuroimaging studies (Calder et al., 2007; Pinsk et al., 2009; Pitcher, Dilks, et al., 2011; Pitcher et al., 2019) have identified a face-selective region in the right anterior STS (raSTS). This, together with the multiple face patches in the macaque STS demonstrates the existence of a cortical pathway projecting down the STS specialized for face perception. Like the pSTS, the amygdala has been strongly implicated in neuroimaging studies of facial expression recognition (Adolphs, 2008; Morris et al., 1996) and lesion studies in humans, and in macaques, report that damage to the amygdala impairs expression recognition (Adolphs et al., 1994; Calder et al., 1996; Hadj-Bouziane et al., 2012). Based on this evidence, a functional connection between the pSTS and amygdala has been proposed in face processing models (Calder & Young, 2005; Haxby et al., 2000). Tracer studies in macaques have also identified a cortical pathway that projects along the STS into the lateral nucleus of the dorsal amygdala (Aggleton et al., 1980; Stefanacci & Amaral, 2000). In human participants we causally mapped the connectivity of this pathway using TMS and fMRI (Pitcher et al., 2017). Participants viewed moving face videos after TMS was delivered over the right pSTS. Results demonstrated that TMS reduced the response to videos of moving faces in the right amygdala, as well as in the face-selective area in the anterior STS. In addition, our visual field mapping study demonstrated that the right amygdala showed no visual field bias (Pitcher et al., 2020) making it more similar to the pSTS, than to the FFA.

While these results suggest that the amygdala is functionally similar to the STS it is important to note that, unlike the pSTS, the amygdala responds equally to moving and static faces (Pitcher et al., 2019). This leaves the connectivity of the amygdala an open question as it exhibits functional properties similar to face areas in both the ventral pathway (FFA), and the lateral pathway (pSTS). It is, of course, highly likely that the amygdala shares functional connections with both the FFA and STS, performing different functional computations using appropriate task related face information from both pathways. This is consistent with a recent study demonstrating that TMS delivered over the right pSTS disrupted resting-state functional connectivity across the entire face network (Handwerker et al., 2020). Crucially, the disruption in connectivity was observed not only between the right pSTS (the stimulation site) and the right amygdala, but also between non stimulated face areas on the ventral surface, namely the right FFA and right amygdala.

Social interactions are not only about understanding the intentions of other people, they also involve a decision about how to respond (even if that decision is to do nothing). Decision making suggests the involvement of brain areas higher in the cortical hierarchy than the STS, notably in the prefrontal cortex. Non-human primate neuroanatomy also identifies reciprocal cortico-cortical pathways that project from the fundus and ventral bank of the anterior STS to ventrolateral prefrontal cortex (Kravitz et al., 2013; Webster et al., 1994). Human and non-human primate studies demonstrate that the prefrontal cortex is involved in attentional control and perceptual decision-making control of category-selective areas in the ventral (e.g. FFA) and dorsal pathways (parietal cortex) (Baldauf & Desimone, 2014; Heekeren et al., 2004; Kim & Shadlen, 1999). The functional connectivity between the prefrontal cortex and the lateral pathway remains comparatively understudied. A recent study large participant study of the face processing network analyzed tractography, rsfMRI and face localizer data from 680 participants (Wang et al., 2020). (Wang et al., 2020) produced results that suggest many future directions. Results demonstrated that the STS and the face area in the inferior frontal gyrus (IFG) were more strongly connected to each other than to any other face area. The authors grouped the STS and IFG into a sub-group of face areas that is specialized for the cognitive processing of moving faces suggesting directions for future studies. In addition, we demonstrated that the right IFG and the right pSTS both exhibit an equal response to faces in both visual fields and that TMS disruption of the right IFG reduces the neural response to faces in the pSTS (Nikel et al., 2022; Pitcher, Sliwinska, et al., 2023).

**Body processing in the third pathway**

The brain's ability to swiftly and effectively process socially relevant visual cues isn't restricted to faces. The third pathway also extracts socially relevant information from the selective processing of body movements. Neurophysiological investigations in non-human primates have been fundamental in establishing this concept, identifying specialized cells within the STS that exhibit selective responses to images of bodies and body parts, such as hands (Desimone et al., 1984; Gross et al., 1972; Wachsmuth et al., 1994). These early findings have since been robustly supported by modern neuroimaging techniques. Subsequent fMRI experiments in both monkeys and humans have pinpointed multiple body-selective areas situated along the STS (Pinsk et al., 2009; Pinsk et al., 2005; Tsao et al., 2003). The profound significance of biological motion in driving neural activity within this region was comprehensively demonstrated by Russ and Leopold (2015). Their study, which involved scanning macaques while they watched videos of conspecifics interacting naturally, revealed that biological motion was the predominant driver of neural responses (Russ & Leopold, 2015). This effect was not isolated; it was observed in a cluster that originated in the early visual cortex and extended along the entire length of the STS. Strikingly, even the classic face-selective patches within the STS exhibited a greater response to biological motion than to static faces, underscoring the dynamic nature of social perception in this cortical area. This highlights the STS's role as an integrator of both identity and action cues.

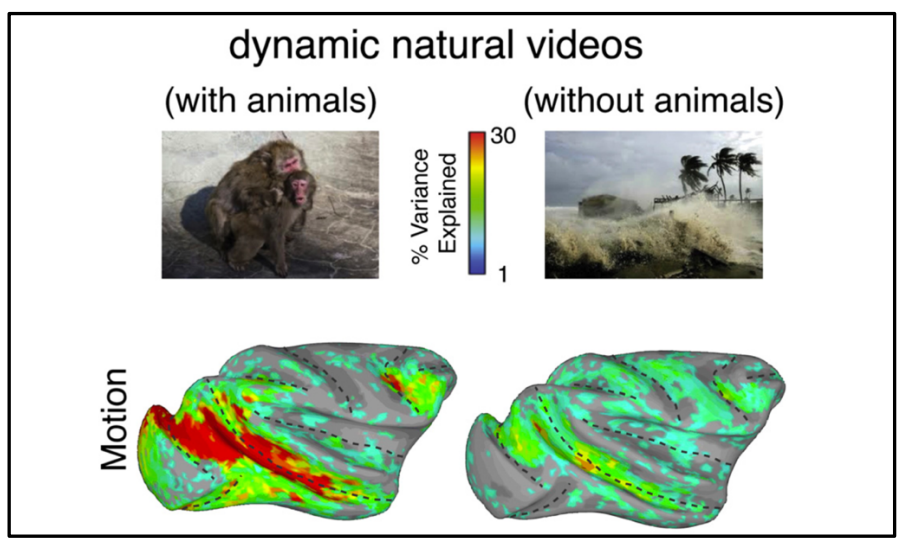

**Figure 7. Left column. Surface maps showing the percent variance explained by the motion (top) and contrast (bottom) feature models to natural movies that contained animals interacting with each other and the environment. Right column. Surface maps showing the percent variance explained by the motion (top) and contrast (bottom) feature models to natural movies that contained no animals. Reduced explained variance for motion in the non-social movies was found in both subjects tested.**

In humans, the neural underpinnings of visual body perception have also been extensively investigated. The STS shows body-selective responses to stimuli like point-light walkers (Grossman & Blake, 2002), moving bodies (Beauchamp et al., 2003), and videos of actors performing physical actions (Saxe et al., 2004). However, the most intensely studied body-selective region in humans is the extrastriate body area (EBA) (Downing et al., 2001; Gandolfo et al., 2024; Kucuk & Pitcher, 2024). The EBA is located slightly inferior and posterior to the STS, residing on the lateral brain surface in Brodmann area 18. This anatomical placement often results in it overlapping with the motion-sensitive area V5/MT (Downing et al., 2007). This spatial proximity strongly suggests that the EBA is also a component of the third visual pathway, heavily implicated in motion processing. Our own evidence supports this, showing that the EBA exhibits a stronger neural response to moving bodies than to static ones (Kucuk et al., 2024; Pitcher et al., 2019). This is further substantiated by clinical observations where a patient with a lesion to the right ventral occipitotemporal cortex showed a normal EBA response to moving bodies (Susilo et al., 2015). Nevertheless, this hypothesis faces a challenge from earlier neuropsychological cases, which reported patients who could perceive biological motion despite having general motion processing impairments (McLeod et al., 1996; Vaina & Gross, 2004; Vaina et al., 1990), suggesting a degree of functional independence that remains an active area of research.

It seems likely that the STS and EBA are functionally distinct components of the third visual pathway specializing in biological perception. The EBA serves a more foundational role as the body recognizer, primarily dedicated to processing the visual form and static posture of the body, establishing the basic categorical identity that a stimulus as a human body part (Peelen & Downing, 2007). In contrast, the STS functions as the social integrator and action interpreter. This region exhibits a profound specialization for dynamic biological motion and goal-directed actions, being the key area for decoding what the body is doing and why (Grossman et al., 2005; Grossman & Blake, 2002; Grossman et al., 2010; Saxe et al., 2004). The STS actively integrates motion information, with evidence showing its classic face-selective patches exhibit a greater response to biological motion than to faces alone (Russ & Leopold, 2015). This suggests that while the EBA focuses on the structural representation of the body, the STS is crucial for extracting the temporally evolving, socially relevant meaning from movement, thereby distinguishing between body as an object and body as a communication tool.

### How soon is now? Timing in the third pathway

Understanding the hierarchy of a visual pathway is dependent not only on mapping the anatomical connectivity but also on mapping the temporal progression of information along the that pathway. Chronometric TMS studies provide a precise experimental method to establish when a targeted brain area is causally engaged in processing the concurrently performed behavioral task. Single or double pulses of TMS can be delivered over the target region at different time points after stimulus onset or after the commencement of behavioral monitoring (Amassian et al., 1993; Pitcher et al., 2012). Plotting the temporal pattern of the induced behavioral impairments reveals when the stimulated area is causally engaged in task performance. Such an approach was used the McGurk effect, a visual multisensory illusion based on the integration of visual and auditory information (Beauchamp et al., 2010). An initial experiment demonstrated that TMS over the right posterior superior temporal sulcus (pSTS) impaired the perception of McGurk stimuli but spared control stimuli. A follow-up chronometric experiment, delivering single-pulse TMS at latencies ranging from -300ms to +300ms relative to stimulus onset, revealed that the pSTS computes the integration of auditory and visual information within a specific temporal window of -100ms to +100ms.

Chronometric TMS is particularly valuable for mapping the temporal dynamics of the face processing network. We employed this method to trace the speed of facial expression recognition across specific nodes (Pitcher et al., 2008). In an initial study, participants performed a delayed match-to-sample task while double-pulse TMS was delivered over the right occipital face area (OFA) and right somatosensory cortex (SC) at various post-stimulus intervals. Results indicated a dissociation in processing timing: TMS over the right OFA disrupted recognition at 60–100ms, whereas TMS over the right SC caused impairment later, at 100–140ms and 130–170ms. A subsequent study (Pitcher, 2014) replicated the early OFA window (60–100ms) and identified a concurrent, though longer, impairment window in the right pSTS (60–140ms).

Collectively, these findings provide causal evidence for the posterior-to-anterior progression of facial expression processing. Notably, the duration of impairment in higher-order areas (pSTS and SC) was nearly double that of the OFA. This difference in the length of the TMS impairment window is consistent with physiological evidence from non-human primates showing that the response profile of neurons in higher cortical regions is longer than the response in earlier cortical regions (Kovacs et al., 1995). Human fMRI evidence also shows that cortical regions in and around the right pSTS show a longer temporal response window to movie clips than regions in early visual cortex (Hasson et al., 2008).

These chronometric findings, however, present a discrepancy with electrophysiological methods. EEG and MEG studies typically report face-selective activity peaking at 100ms and 170ms (Bentin et al., 1996; Liu et al., 2002) and sometimes peaking as late as 500ms (Yan et al., 2025), latencies that are notably later than the critical TMS windows observed in the OFA and pSTS. This divergence may arise because EEG/MEG record the summation of activity across multiple areas, whereas TMS targets focal disruption. Alternatively, the discrepancy may reflect the mechanism of stimulation: TMS may be most effective when disrupting the buildup of neuronal activity rather than its peak. This is consistent macaque data showing a face-selective local field potential (LFP) response peaking at 150ms in the temporal lobe (Afraz et al., 2006). Crucially, the greatest impairment in face detection accuracy was observed after microstimulation delivered at 50-100ms, not the 100-150ms stimulation that directly preceded the LFP peak at 150ms. This suggests a difference between methods. Namely, disrupting task performance (with TMS or microstimulation) is most effective when the neural activity that supports task performance is building. By contrast, the peak of neural activity (in MEG or LFP) represents neural activity that has already occurred, and so effective disruption is less effective or too late (Pitcher, 2024).

**<u>Right or left? The laterality of social cognition in the third visual pathway</u>**

Empirical data collected using different experimental methods have demonstrated that face recognition is preferentially processed in the right hemisphere (Barton et al., 2002; Kanwisher et al., 1997; Landis et al., 1986; Pitcher et al., 2007; Young et al., 1985; Yovel et al., 2003). This right hemisphere dominance for face recognition is also present in the STS as shown by neuroimaging (Kucuk et al., 2024; Pitcher, Dilks, et al., 2011) and TMS studies (Sliwinska, Elson, et al., 2020; Sliwinska & Pitcher, 2018). However social perception is a much more complicated process than simply recognising facial expressions. This was elegantly demonstrated in a recent fMRI study in which participants watched 250 3-second videos of two people performing a range of different social interactions (e.g., two people doing Karate or two people reading a map). The social scenes were then curated to identify the visual features (McMahon et al., 2023). These include low-level features (e.g., contrast and motion energy), mid-level features (e.g., physical distance between the actors and their direction of attention) and high-level features that support social understanding (e.g., the nature and valence of the interaction). Results were consistent with the cortical hierarchy predicted by the visual third pathway. Motion was preferentially processed in V1 and V5/MT, mid-level features corresponding to scene geometry were preferentially processed in the EBA and LO, and high-level features corresponding to the nature and intensity of the interaction were preferentially processed in the STS. Crucially, the videos that depicted greater levels of communication (e.g., more social interaction and higher valence) were preferentially processed in the right hemisphere.

However, it is also important to note that the left hemisphere still exhibits a visual response to faces and other types of biological motion. This was clearly demonstrated in an fMRI study that examined the response to a range of tasks including faces, bodies, biological motion, spoken language and theory of mind (Deen et al., 2015). The posterior STS (bilaterally, but right dominant) responded to transient social inputs like motion and faces. But this bilateral changed when moving anteriorly along the length of the STS. Namely, the anterior STS (particularly in the left hemisphere) becomes increasingly selective for speech and intelligible language. This is also consistent with Anatomical studies show the planum temporale, a brain area adjacent to the STS that processes speech is larger in the left hemisphere (Geschwind & Levitsky, 1968). This is consistent with a model of laterality in the STS suggesting that while the right STS extracts and interprets the visual cues that underpin social cognition. While the left STS integrates these signals with speech into a social communicative framework (Redcay, 2008).

## Conclusion

Charles Darwin claimed that there are two types of scientist: "lumpers" and "splitters". Lumpers seek the fundamental similarities that unify the functionality of a biological system while splitters look to further quantify the finer and finer differences between the dissociable components in a biological system. The theoretical strength of visual pathway models (Milner & Goodale, 1986; Ungerleider & Mishkin, 1982) is that they allow us to both lumpers and splitters. Physiological and neuroimaging research can be used to characterize the response of neurons and cortical areas to different visual stimuli to provide a precise spatial map of the brain. Visual pathways group these category-selective responses together based on anatomical and functional connectivity. Visual pathway models create a common cognitive framework that enables scientists to integrate empirical data using a broad range of experimental methods. This facilitates understanding between those who study the brain at the behavioral, cognitive and neural levels. The third visual pathway (Pitcher & Ungerleider, 2021) represented our effort to update prior models of primate visual cortex by incorporating the extensive empirical research produced in the decades since the original two-pathway model, thereby providing a more comprehensive understanding of the neural mechanisms underlying social perception.

## Acknowledgements

This work was supported by Leverhulme Trust Project Grant RPG-2024-389 (D.P.).

**Pitcher, D.** The Third Visual Pathway for Social Perception.
*The Oxford Handbook of Face Perception (2nd Edition). Oxford University Press.*

Grossmann, T. (2021). Developmental Origins of the Pathway for Social Perception. *Trends Cogn Sci*. https://doi.org/10.1016/j.tics.2021.03.003

Gschwind, M., Pourtois, G., Schwartz, S., de Ville, D. V., & Vuilleumier, P. (2012). White-Matter Connectivity between Face-Responsive Regions in the Human Brain. *Cerebral Cortex*, *22*(7), 1564–1576. https://doi.org/10.1093/cercor/bhr226

Hadj-Bouziane, F., Liu, N., Bell, A. H., Gothard, K. M., Luh, W. M., Tootell, R. B., Murray, E. A., & Ungerleider, L. G. (2012). Amygdala lesions disrupt modulation of functional MRI activity evoked by facial expression in the monkey inferior temporal cortex. *Proc Natl Acad Sci U S A*, *109*(52), E3640–3648. https://doi.org/10.1073/pnas.1218406109

Handwerker, D. A., Ianni, G., Gutierrez, B., Roopchansingh, V., Gonzalez-Castillo, J., Chen, G., Bandettini, P. A., Ungerleider, L. G., & Pitcher, D. (2020). Theta-burst TMS to the posterior superior temporal sulcus decreases resting-state fMRI connectivity across the face processing network. *Netw Neurosci*, *4*(3), 746–760. https://doi.org/10.1162/netn_a_00145

Hasson, U., Yang, E., Vallines, I., Heeger, D. J., & Rubin, N. (2008). A hierarchy of temporal receptive windows in human cortex. *J Neurosci*, *28*(10), 2539–2550. https://doi.org/10.1523/JNEUROSCI.5487-07.2008

Haxby, J. V., Hoffman, E. A., & Gobbini, M. I. (2000). The distributed human neural system for face perception. *Trends Cogn Sci*, *4*(6), 223–233.

Heekeren, H. R., Marrett, S., Bandettini, P. A., & Ungerleider, L. G. (2004). A general mechanism for perceptual decision-making in the human brain. *Nature*, *431*(7010), 859–862. https://doi.org/10.1038/nature02966

Hein, G., & Knight, R. T. (2008). Superior temporal sulcus--It's my area: or is it? *J Cogn Neurosci*, *20*(12), 2125–2136. https://doi.org/10.1162/jocn.2008.20148

Hemond, C. C., Kanwisher, N. G., & Op de Beeck, H. P. (2007). A preference for contralateral stimuli in human object- and face-selective cortex. *PLoS One*, *2*(6), e574. https://doi.org/10.1371/journal.pone.0000574

Huk, A. C., Dougherty, R. F., & Heeger, D. J. (2002). Retinotopy and functional subdivision of human areas MT and MST. *Journal of Neuroscience*, *22*(16), 7195–7205.

Im, E. J., Shirahatti, A., & Isik, L. (2025). Early Neural Development of Social Interaction Perception: Evidence from Voxel-Wise Encoding in Young Children and Adults. *J Neurosci*, *45*(1). https://doi.org/10.1523/JNEUROSCI.2284-23.2024

Isik, L., Koldewyn, K., Beeler, D., & Kanwisher, N. (2017). Perceiving social interactions in the posterior superior temporal sulcus. *Proc Natl Acad Sci U S A*, *114*(43), E9145–E9152. https://doi.org/10.1073/pnas.1714471114

Kanwisher, N., McDermott, J., & Chun, M. M. (1997). The fusiform face area: a module in human extrastriate cortex specialized for face perception. *J Neurosci*, *17*(11), 4302–4311.

Kay, K. N., Weiner, K. S., & Grill-Spector, K. (2015). Attention reduces spatial uncertainty in human ventral temporal cortex. *Curr Biol*, *25*(5), 595–600. https://doi.org/10.1016/j.cub.2014.12.050

Kilner, J. M. (2011). More than one pathway to action understanding. *Trends Cogn Sci*, *15*(8), 352–357. https://doi.org/10.1016/j.tics.2011.06.005

Kim, J. N., & Shadlen, M. N. (1999). Neural correlates of a decision in the dorsolateral prefrontal cortex of the macaque. *Nat Neurosci*, *2*(2), 176–185. https://doi.org/10.1038/5739

Kornblith, S., Cheng, X., Ohayon, S., & Tsao, D. Y. (2013). A network for scene processing in the macaque temporal lobe. *Neuron*, *79*(4), 766–781. https://doi.org/10.1016/j.neuron.2013.06.015

Kosakowski, H. L., Cohen, M. A., Takahashi, A., Keil, B., Kanwisher, N., & Saxe, R. (2022). Selective responses to faces, scenes, and bodies in the ventral visual pathway of infants. *Curr Biol*, *32*(2), 265–274 e265. https://doi.org/10.1016/j.cub.2021.10.064
28